\documentclass[final,1p,times]{elsarticle} % do not change
\usepackage{graphicx} % do not change
\usepackage{amssymb} % do not change
\usepackage{amsthm} % do not change
\usepackage{lineno} % do not change

\usepackage{bm}
\usepackage{amsmath}
\def\tfrac#1#2{{\textstyle\frac#1#2}}

\journal{Nuclear Physics A} % do not change
\begin{document} % do not change

\begin{frontmatter} % do not change

%% QM09Author: please enter your  
%% Title, author and address info here; please do not use footnotes

% Your Title - please modify
\title{Gluon Bremsstrahlung in Weakly-Coupled Plasmas}

% Principle author, and co-authors - please modify
\author{Peter Arnold}

% Address - please modify
% note that if you have authors from several institutions, we recommend
% labelling these [a], [b], [c] etc.
\address{University of Virginia, % label [a]
P.O. Box 400714,
Charlottesville, VA, 22904-4714, USA}

\begin{abstract} % do not change
%% Text of abstract goes here - please modify
I report on some theoretical progress concerning the
calculation of gluon bremsstrahlung for very
high energy particles crossing a weakly-coupled quark-gluon plasma.
(i) I advertise that two of the several formalisms used to study this
problem, the BDMPS-Zakharov formalism and the AMY formalism (the latter
used only for infinite, uniform media), can be made equivalent when
appropriately formulated.  (ii) A standard technique to simplify
calculations is to expand in inverse powers
of logarithms $\ln(E/T)$.
I give an example where such expansions are found to work well
for $\omega/T \gtrsim 10$
where $\omega$ is the bremsstrahlung gluon energy.
(iii) Finally, I report on perturbative
calculations of $\hat q$.
\end{abstract} % do not change

\end{frontmatter} % do not change

%% QM09: we keep linenumbers at least for initial version
%\linenumbers % do not change

%% start of main text - please modify. Below is a sub-set (single section) 
%% of an earlier proceedings that show how one can handle references 
%% and figures etc.
%%\section{}\label{}

\def\md{m_{\rm D}}

\section{Equivalence of BDMPS-Zakharov and AMY}

The calculation of gluon bremsstrahlung from high-energy particles
traversing a quark-gluon plasma is complicated by the
Landau-Pomeranchuk Migdal (LPM) effect:
at high energy,
the formation length for bremsstrahlung (the distance over which
gluon or photon emission is coherent) becomes longer than the mean
free path between scatterings.  As a result, bremsstrahlung from
successive scatterings cannot be treated as independent.
For QED, Migdal solved this problem in 1956.
The basic formalism for treating the LPM effect in QCD
was worked out by Baier, Dokshitzer,
Mueller, Peign\'e and Schiff (BDMPS) and Zakharov in
1996-1998 \cite{BSZ}.
There have been a number of variations on attacking the problem
since then, but the one I want to focus on here is that of
myself, Moore, and Yaffe (AMY) \cite{AMYsansra}.
The AMY formalism has
been used for the calculation of transport coefficients, such
as shear and bulk viscosity, to leading-order in $\alpha_{\rm s}$
\cite {transport1,transport2}.
The AMY formalism was derived independently because, at the time, we
did not understand the BDMPS-Zakharov formalism and could not
see how to overcome some of its limitations.

One of the limitations of the AMY formalism was that it was formulated
only for the case of a uniform, infinite-volume system, which means
systems which are approximately uniform over one formation length
(and do not change over the corresponding formation time).
BDMPS-Zakharov,
in contrast, could handle the non-uniform, time-dependent
case as well, such as a system
whose size is smaller than the formation length.  On the downside,
BDMPS-Zakharov implicitly treated plasma particles as static scatterers.
Furthermore, their results were normalized in terms of the gluon mean
free path $\lambda$.  That's a problem for small coupling expansions
because $\lambda$ is {\it zero}
in (hard-thermal-loop resummed) perturbation theory!
A popular model for $\lambda$, often used as an example by BDMPS, is
perturbative Coulomb scattering cut off by Debye screening, with the form
\begin {equation}
  \lambda^{-1} = \int d^2 q_\perp \frac{\# g^4 n}{(q_\perp^2 + \md^2)^2}
  ,
\label{eq:lambda1}
\end {equation}
where $q_\perp$ is the transverse momentum transfer, $n{\sim}T^3$ is
the density of plasma particles, and $\md{\sim}gT$ is the Debye mass.
The Debye effect screens electric fields, but it does not screen
nearly-static components of magnetic fields.  As a result,
magnetic scattering can take place at much smaller $q_\perp$,
and the actual formula representing perturbative scattering with
screening effects is
\begin {equation}
  \lambda^{-1} = \int d^2 q_\perp
     \frac{\# g^2 \md^2}{q_\perp^2 (q_\perp^2 + \md^2)} .
\end {equation}
This is infrared log divergent, giving $\lambda=0$.

The result I have to report \cite{simple} is that, with minor
tweaking of the way BDMPS-Zakharov write their formulas, one
can eliminate all the above issues and then easily show that
BDMPS-Zakharov reproduces AMY in the uniform, infinite medium
limit.  So, there is now a BDMPS-Zakharov formalism that handles
non-static scatterers, and equivalently there is now a version of
AMY for non-uniform media.  See Ref. \cite{simple} (section II.B
and the appendix) for details.  The basic tweak to BDMPS-Zakharov
is to write their formulas more generally in terms of the rate
$\Gamma_{\rm el}$
for elastic scattering off of the medium instead of as density $n$
times the corresponding
cross-section $\sigma$, and to avoid normalizing quantities by $\lambda$.

\section{Large logarithm approximation}

To fully evaluate the bremsstrahlung rate to leading order in
$\alpha_{\rm s}$ requires somewhat complicated numerics, even in the case of
uniform media. There is a great deal of simplification if one
makes the additional approximation that the logarithm $\ln(E/T)$
of energy is large [while still treating $\alpha_{\rm s} \ln(E/T) \ll 1]$.
With this large logarithm approximation, it is possible to get analytic
results.  To make practical use of a large logarithm expansion, however,
one needs a next-to-leading log (NLL) result.  Leading log cannot tell
the difference, for example, between
\begin {equation}
   \ln\left(\frac{E}{T}\right)
   \qquad
   \mbox{and}
   \qquad
   \ln\left(\frac{E}{4\pi^2T}\right)
   = \ln\left(\frac{E}{T}\right) + O(1) .
\end {equation}
But these two logarithms are very different for realistic values of
$E/T$.

\begin{figure}[ht]
\centering
\includegraphics[scale=0.40]{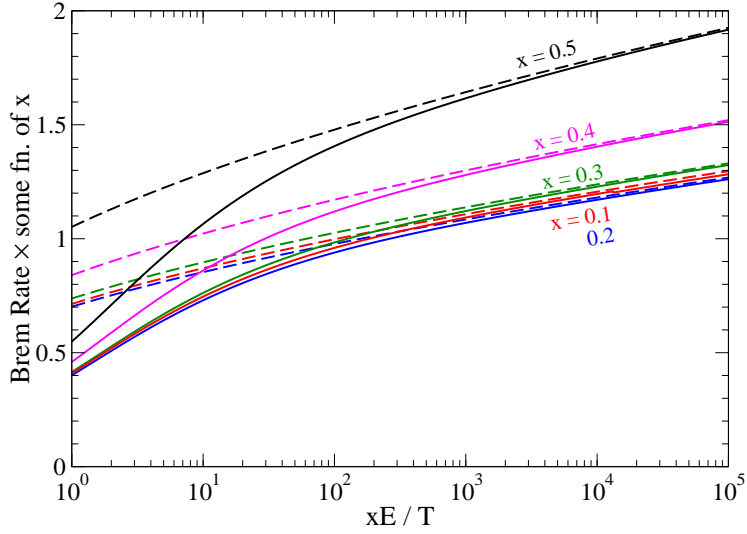}
\caption[]{$g{\to}gg$ bremsstrahlung rate in an infinite, uniform
  medium.
  Solid line is a full
  calculation to leading order in $\alpha_{\rm s}$.
  Making the further approximation that $\ln(xE/T)$ is large,
  and working to NLL order, gives the dashed line.
  See Ref.\ \cite{ArnoldDogan} for normalization of vertical axis.
}
\label{fig:ArnoldDogan}
\end{figure}

But, even if one has a NLL result, is it any use in practical
situations, or is $\log(E/T)$ never large enough?  A test is shown in
Fig.\ \ref{fig:ArnoldDogan}, which gives results from Ref.\
\cite{ArnoldDogan}.  The solid lines are results from a full
leading-order evaluation of the $g{\to}gg$ bremsstrahlung rate
(equivalent to the sort of calculations first performed by Jeon and
Moore \cite{JeonMoore} based on the AMY formalism), as a function of the
bremsstrahlung gluon energy $\omega = x E$ divided by $T$.  The dashed
lines show the result of the NLL calculation of Ref.\
\cite{ArnoldDogan}.  The horizontal axes goes up to the very large, absurdly
unrealistic value $\omega/T = 10^5$ just to verify that the
curves do approach each other as $\ln(E/T) \to \infty$ (for fixed $x$),
as they should.  The conclusion to take away from this plot is that,
in the context of a weakly-coupled plasma, the large logarithm
expansion to NLL is good to $\lesssim 20\%$ for $\omega \gtrsim 10 T$,
which is much better than one might have feared.
For $\omega \sim T$, the NLL results is off from the full
small-coupling result by roughly a factor of two.

\section{$\hat q$ in weakly-coupled plasmas}

Finally, I want to wrap up this potpourri by discussing the value
of the jet broadening parameter $\hat q$ in the limit of weak coupling.
$\hat q$ is defined as the averaged squared transverse momentum transfer
$Q_\perp^2$ per unit length to
a high-energy particle traversing the plasma,
so that $Q_\perp^2 = \hat q L$.  It's of relevance
to bremsstrahlung because the formation time depends on the
collinearity of the bremsstrahlung gluon with the particle that
emits it, and the degree of collinearity in turn depends on how much the
particles are randomly deflected during the bremsstrahlung process.

\begin{figure}[t]
\centering
\includegraphics[scale=0.40]{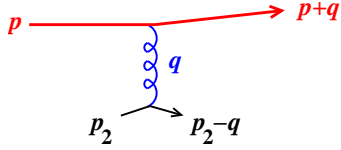}
\caption[]{
  Elastic scattering of a high-energy particle ($p$) off of a
  plasma particle ($p_2$).
}
\label{fig:qhat}
\end{figure}

The squared transverse momentum transfer per unit length is simply
\begin {equation}
   \hat q = \int d^2 q_\perp \> \frac{d\Gamma_{\rm el}}{d^2q_\perp} \,
            q_\perp^2 ,
\label {eq:qhat}
\end {equation}
where $\Gamma_{\rm el}$ is the rate for elastic scattering from
plasma particles, as in Fig.\ \ref{fig:qhat},
and $q_\perp$ is the transverse momentum transfer in a single collision.
When evaluated at leading-order in $\alpha_{\rm s}$ (and ignoring the
running of $\alpha_{\rm s}$), this formula is UV log divergent.
For the bremsstrahlung problem, however, what one actually needs is
the result $\hat q(\Lambda)$ obtained by introducing a UV cut-off
$\Lambda$ on the $q_\perp$ integration \cite{BDMPS3,ArnoldXiao,simon}.

The calculation of the $d\Gamma_{\rm el}/d^2 q_\perp$ needed in
(\ref{eq:qhat}), based on the leading-order process of Fig.\ \ref{fig:qhat},
has the form
\begin {equation}
   \frac{d\Gamma_{\rm el}}{d^2 q_\perp}
   \sim \int d q_z \int d^3 p_2 \>
      \frac{d\sigma_{\rm el}}{d^3 q}
      f({\bm p}_2) [1 \pm f({\bm p}_2 - {\bm q})] ,
\label {eq:dgam}
\end {equation}
where $f({\bm p})$ is the Bose or Fermi distribution for finding a
particle with momentum ${\bm p}$ in the plasma.  The $1\pm f$ factor is
a Bose enhancement or Pauli blocking factor for the final state of
the plasma particle.  Most perturbative calculations of $\hat q$ have
made additional simplifications to this formula.  Some implicitly ignore the
final-state $1{\pm}f$ factor, which is ignorable only when the dominant
momentum transfers $q$ are large compared to $T$, and even then only
to leading-log order.  Others replace
$1\pm f({\bm p}_2-{\bm q})$ by $1\pm f({\bm p}_2)$, valid if
the dominant $q$ are small compared to $T$.
To my knowledge, the result has only recently been evaluated using
the full form (\ref{eq:dgam}) \cite{ArnoldXiao,simon}.
For $\Lambda \gg T$ [but $\alpha_{\rm s} \ln(\Lambda/T) \ll 1]$,
an analytic small-coupling result for UV-regulated $\hat q$ is
given in Ref.\ \cite{ArnoldXiao}.  As an example, for a
purely gluonic plasma, it is%
\footnote{
  For comparison, if one made the $q \ll T$ approximation of
  replacing $1+f({\bm p}_2-{\bm q})$ by $1+f({\bm p}_2)$ in
  (\ref{eq:dgam}), the result would be
  $
    \hat q_{\rm g}(\Lambda) =
    \left[ \zeta(2) \ln \frac{\Lambda}{\md}\right]
    \frac{9 g^4 T^3}{\pi^3}
  $.
  If one instead made the $q \gg T$ approximation of dropping
  the $1{+}f$ term in (\ref{eq:dgam}), the result would be
  $
    \hat q_{\rm g}(\Lambda) =
    \left[ \zeta(3) \ln \frac{\Lambda}{\md}\right]
    \frac{9 g^4 T^3}{\pi^3}
  $.
  A rough approximation one sees in the literature is to
  (i) replace the final factor of $q_\perp^2$ in (\ref{eq:qhat}) by
  $\md^2$, which artificially eliminates the UV log divergence
  and gives $\hat q \simeq \md^2/\lambda$,
  and (ii) use the model form (\ref{eq:lambda1}) for the gluon
  mean free path $\lambda$.
  This gives
  $
    \hat q_{\rm g} \simeq
    \left[ \tfrac12 \, \zeta(3) \right]
    \frac{9 g^4 T^3}{\pi^3}
  $
  in the purely gluonic case, for which
  $\# = 9 \zeta(3)/2\pi^4$ in (\ref{eq:lambda1}).
}
\begin {subequations}
\label{eq:qhatresult}
\begin {equation}
   \hat q_{\rm g}(\Lambda) =
   \left[ \zeta(3) \ln \frac{\Lambda}{c T}
          + \zeta(2) \ln\frac{c T}{\md}
          - \sigma_+
   \right]
   \frac{9 g^4 T^3}{\pi^3} ,
\end {equation}
where
\begin {equation}
   c \equiv 2 \exp\left( \tfrac12 - \gamma_{\rm E}\right) ,
\end {equation}
\begin {equation}
  \sigma_+ \equiv \sum_{k=1}^\infty \frac{\ln[(k-1)!]}{k^3}
  = 0.386043817389949...,
\end {equation}
\end {subequations}
and $\md = g T$ is the Debye mass.
The constant
$\sigma_+$ can be related to certain generalizations of the Riemann $\zeta$
function.
I think the expression (\ref{eq:qhatresult})
for $\hat q$ is fun and interesting.
But if you are of a more practical bent,
you could have instead just done the integral
(\ref{eq:dgam}) numerically.

A presentation of $\hat q$ to leading order in $\alpha_{\rm s}$ requires
an important warning: Corrections which are formally higher-order in
coupling, of order $\md/T = O(g)$, have been analyzed by Caron-Huot
\cite{simon} and
are of order 100\% for realistic couplings.

%% end of main text

\section*{Acknowledgments} % please check/modify, comment out or delete if not needed
This work was supported, in part, by the U.S. Department
of Energy under Grant No.~DE-FG02-97ER41027.

 % do not change 
\end{document}